\documentclass[twocolumn,showpacs,preprintnumbers,amsmath,amssymb]{revtex4}
\usepackage{epsfig}
\begin{document}

\title{Symmetry of Magnetic Quantum Tunneling in Single Molecule
Magnet Mn$_{12}$-acetate}
\author{E. del Barco$^1$, A. D. Kent$^1$, E. M. Rumberger$^2$, D. N. Hendrickson$^2$ and G. Christou$^3$}
\affiliation{$^1$Department of Physics, New York University, 4 Washington Place. New York, NY 10003}
\affiliation{$^2$Department of Chemistry and Biochemistry, University of California San Diego - La Jolla, CA
92093-0358} \affiliation{$^3$Department of Chemistry, University of Florida ,Gainsville, FL 32611-7200}
\date{Received 9 April 2003: published 24 July 2003}

\begin{abstract}
The symmetry of magnetic quantum tunneling has been studied in the prototype single molecule magnet
Mn$_{12}$-acetate using a micro-Hall effect magnetometer and superconducting high field vector magnet system. An
average crystal fourfold symmetry is shown to be due to local molecular environments of twofold symmetry that are
rotated by 90$^o$ with respect to one another, confirming that disorder which lowers the molecule symmetry is at
important to magnetic quantum tunneling. We have studied a subset of these lower (twofold) site symmetry molecules
and present evidence for a Berry phase effect consistent with a local twofold symmetry.
\end{abstract}
 \maketitle

Magnetic quantum tunneling (MQT) in molecules is a very active area of research, starting with the initial
observations of resonant quantum tunneling in single molecule magnet (SMM) Mn$_{12}$-acetate
\cite{Friedman,Thomas}. Quantum tunneling enables the creation of superposition states necessary for quantum
computing and is also a mechanism for the loss of information in classical magnetic information storage. Condensed
matter systems like these \cite{Vion,Nakamura}, with discrete energy levels coupled to the environment in the
solid, are also important to understanding the border between quantum and classical physics, and the decoherence
of quantum systems \cite{Prokofev}. The symmetry of the molecule is the first step in the understanding its energy
level structure and quantum tunneling. For example, the symmetry determines the allowed terms in the spin
Hamiltonian and, in particular, the form of the transverse interactions that lead to tunneling. However, in
Mn$_{12}$ as well as other SMMs MQT does not reflect the molecule site symmetry and the origin of tunneling has
been an open question
\cite{Hernandez1,Sangregorio,Barra,Wernsdorfer1,Bokacheva,Mirabeau,Hill1,delBarco1,Chudnovsky,Cornia,Mertes,Hernandez2,delBarco2,Hill2}.

Mn$_{12}$-acetate [Mn$_{12}$O$_{12}$(CH$_{3}$COO)$_{16}$(H$_{2}$O)$_{4}$]-2CH$_{3}$COOH-4H$_{2}$O consists of a
core of twelve manganese atoms with a ground state spin of 10. The spin Hamiltonian of an ideal (S$_4$-symmetry)
Mn$_{12}$-acetate molecule is:
\begin{equation}
\label{eq.1}{\cal {H}}=-DS_z^2-BS_z^4+C(S_+^4+S_-^4)-g\mu _B{\bf
{H}\cdot {S}}\;.
\end{equation}
The first two terms represent the uniaxial magnetic anisotropy of the molecule ($D >$ 0 and $B >$ 0). The
parameters, $D=$ 0.548 K, $B=1.1\times 10^{-3}$ K and $C=3\times 10^{-5}$ K, have been determined by inelastic
neutron spectroscopy \cite{Mirabeau} and EPR \cite{Hill1} experiments. There are thus 2$S$+1 = 21 allowed
projections of the spin on the z-axis, with up and down projections along the $z$-axis separated by an anisotropy
barrier of $\sim DS^2+BS^4$ ($\sim$ 66 K). An applied longitudinal magnetic field, $H_z$, shifts the energy levels
favoring the projections of the magnetization parallel to the field direction. There are values of the $z$-axis
field (resonance fields) for which the levels $m$ and $m'$ with antiparallel projections on the $z$-axis are
nearly degenerate, $H_k\sim kD/g\mu_B$, $k=m+m'$, ($k$0.44T) and transitions across the energy barrier can occur.
Interactions that break the axial symmetry and mix the levels $m$ and $m'$ lead to an energy difference between
symmetric and antisymmetric linear combinations of spin-projections known as the tunnel splitting. The lowest
order transverse term allowed by the tetragonal symmetry is fourth order (i.e., the third term in eq. 1) and would
lead to the tunneling selection rule $m-m'=4i$, with $i$ an integer.

This is not observed experimentally. In addition, recent experiments show that there is a distribution of
tunneling splittings in Mn$_{12}$ crystals \cite{Mertes,Hernandez2} associated with disorder.  Two distinct models
of disorder have been proposed. Chudnovsky and Garanin \cite{Chudnovsky} proposed that random line dislocations in
a crystal lead, via magnetoelastic interactions, to a lower molecule symmetry and a broad distribution of
tunneling rates. Subsequent magnetic relaxation experiments indeed showed the existence of a broad distribution of
tunneling rates and were analyzed in terms of this model \cite{Mertes,Hernandez2}. In contrast, Cornia et al.
\cite{Cornia} suggested, based on detailed x-ray analysis, that variations in the position of the two
hydrogen-bonded acetic acid molecules surrounding the Mn$_{12}$ clusters lead to a discrete set of isomers with
lower symmetry than tetragonal. Our recent magnetic relaxation experiments in a longitudinal (easy axis) magnetic
field are consistent with solvent disorder \cite{delBarco2}. However, direct evidence of lower symmetry molecular
environments in MQT experiments was lacking.

In this letter we present studies of the symmetry of the magnetic response in deuterated Mn$_{12}$ single crystals
in the pure quantum regime ($T=$ 0.6 K), in which relaxation is by MQT without thermal activation
\cite{Bokacheva,Note1}. An average fourfold symmetry is unambiguously shown to be due to local molecular
environments of twofold symmetry that are rotated by 90 degrees with respect to one another, in accord with the
model proposed by Cornia \cite{Cornia}. Further, we have studied a subset of these lower site symmetry molecules
and present evidence for a Berry phase effect consistent with a local lower (twofold) symmetry.

The magnetization component parallel to the axial direction ($z$-axis) of a Mn$_{12}$-acetate single crystal was
measured using a high sensitivity micro-Hall effect magnetometer in a low temperature Helium 3 system \cite{Kent}.
A single crystal was placed with one of its faces parallel to the plane of the sensor while a high field
superconducting vector field magnet was used to apply magnetic fields at arbitrary directions with respect to the
crystallographic axes of the sample. To study MQT rates we sweep the applied $z$-axis field at a constant rate
$(\alpha=dH_z/dt)$ through a resonance and measure the change in sample magnetization, $M_{before}-M_{after}$
\cite{Wernsdorfer2}. The normalized magnetization change $(M_{before}-M_{after})/(M_{before}-M_{eq})$, where
$M_{eq}$ is the equilibrium magnetization, is the MQT probability, $P$. For a monodisperse system of molecules,
this probability is related to the quantum splitting of the resonance, $\Delta_k$, through the Landau-Zener
formula $P_{LZ}=1-exp(-\pi\Delta^2_k/2\nu_0\alpha)$, where $\nu_0=g\mu_B(2S-k)$ and $\nu_0\alpha$ is the energy
sweep rate. When there is a distribution of quantum splittings in a crystal this situation changes. Now the MQT
probability depends on the distribution of tunnel splittings of the molecules in the initial state $m=10$ (the
metastable state) prior to crossing the resonance. Therefore, it is possible to study different parts of the
distribution (such as molecules with either the largest or smallest tunnel splittings) by appropriate preparation
of the initial magnetization state, as will be described below.

In our first experiments, two different magnetization states were studied. In the first case a) $M_{initial}=M_s$
and we study relaxation with the field applied along the -$z$ direction ($M_{eq}=-M_s$). The initial state
corresponds to all molecules in the state $m$ = 10 and the entire distribution of quantum splittings (100$\%$)
contributes to the relaxation as the field is increased along the -$z$ direction. Note that for the smallest
observable resonances the molecules with the largest tunnel splittings within this distribution will make the
dominant contribution. In the second case, b) we start with $M_{initial}=-0.4M_s$, having allowed 70 $\%$ of the
molecules to relax prior to the experiment by crossing resonance $k=6$ (in the absence of an applied transverse
field). We are thus examining 30$\%$ of the molecules with smallest tunneling splittings. After we prepare these
magnetic states we sweep the longitudinal magnetic field at a constant sweep rate ($\alpha=$ 0.4 T/min) from 0 T
to about 5 T, while a transverse field of 0.4 T is applied at an angle, $\phi$, with respect to one of the faces
of the crystal. We repeat this procedure for angles from $\phi$ = 0$^o$ to $\phi=360^o$.
\begin{figure}
\begin{center}\includegraphics[width=7cm]{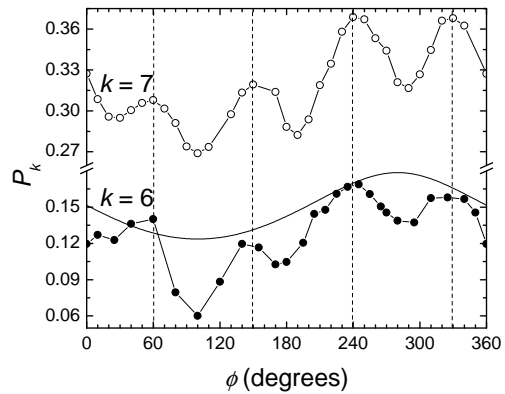}
\vspace{-4 mm}\caption{MQT relaxation probability versus the angle, $\phi$ , between the applied transverse
magnetic field and one of the faces of the crystal for resonance $k$ = 6 (with the whole crystal) and $k$ = 7 (for
30$\%$ of the molecules with smallest tunnel splittings). The solid line represents the effect of misalignment
between the longitudinal field and the easy axis of the crystal.}\vspace{-8 mm}
\end{center}
\end{figure}

In fig. 1 we show the angular dependence of the MQT relaxation probability for both initial configurations.
Resonance $k$ = 6 is shown for case (a) and $k$ = 7 for case (b). In both cases the result is a four-fold pattern
of maxima in the MQT probability at $\phi=60^o$, 150$^o$, 240$^o$ and 330$^o$. These maxima correspond to an
enhancement of the relaxation rates when the transverse field is applied in these directions. There is also a
one-fold contribution due to a small misalignment between the applied field and the $z$-axis of the crystal which
is represented by a continuous line in the figure. It is clear that both initial magnetization states show the
same behavior for the probability versus angle of the transverse field. Since these initial configurations
represent two different parts of the tunnel splittings distribution (high and low ends of the distribution), this
indicates that the four-fold symmetry of the MQT probability is a property of a significant fraction of the
molecules in the crystal. This four-fold rotation pattern is qualitatively consistent with the fourth order
transverse anisotropy term in the Hamiltonian of eq. (1). Note that this term leads to two hard and two medium
magnetic axes in the x-y plane. For example, for positive $C$ the $x$ and $y$ ($\phi=n\pi/2$, with $n=$ 0,1,2,3)
axes are hard and $x$ = $\pm y$ $(\phi=(2n+1)\pi/2)$ are medium axes. A magnetic field applied parallel to a
medium magnetic axis produces a larger tunnel splitting than the same field applied along a hard axis
\cite{Note2}. However, the value of the parameter $C$ found by EPR spectroscopy, $\sim 3\times 10^{-5}$ K, would
produce a change in tunneling probability of only $\sim 15\%$, much smaller than the results observed in the lower
curve in fig. 1.

In order to determine the origin of this four-fold symmetry and whether it is intrinsic to the Mn$_{12}$ molecule
we have conducted the following experiment. We have selected a fraction of molecules in the crystal by applying a
{\it selection transverse field} (STF), $H_{STF}$ = 0.6 T, aligned with one of the maxima in fig. 1, i.e. parallel
to a medium magnetic axis, while crossing resonance $k$, with the longitudinal field swept along $+z$. Then we study this fraction of molecules: we measure
the magnetization versus longitudinal with the field swept in the {\it opposite direction}, along $-z$, in the presence of a transverse field of 0.3 T applied at different
angles, $\phi$, as was done in Fig. 1.  First, the STF was set at $\phi_{select}=60^o$ and in separate experiments both 50$\%$ (selection with $k$ = 6) and 10$\%$ (selection with
$k$ = 5) of molecules with the largest tunnel splitting were studied for this angle of the STF. The same procedure was repeated with the STF in a direction orthogonal to
the first case ($\phi_{select}=150^o$), with 50$\%$ of molecules with the largest tunnel splittings. The behavior
of the MQT probability versus the angle, $\phi$, is shown in fig. 2 for resonance $k$ = 6 and for the three
initial magnetization states described above. The results in all the cases show a {\it two-fold rotation pattern}
of the MQT probability, in clear contrast with the four-fold rotation pattern observed when there was no
transverse field used to select the initial state of the sample. When the STF was applied at $\phi_{select}$ =
60$^o$ (left figure), the results show only two maxima at 60$^o$ and 240$^o$. On the other hand, for the selection
at $\phi_{select}=150^o$ the probability shows maxima at 150$^o$ and 330$^o$. In the left polar plot in fig. 2 we
show the results for both angle selections, $\phi_{select}$ = 60$^o$ (solid circles) $\phi_{select}=150^o$ (open
circles), where the effect of the field misalignment has been corrected.
\begin{figure}
\begin{center}\includegraphics[width=8.6cm]{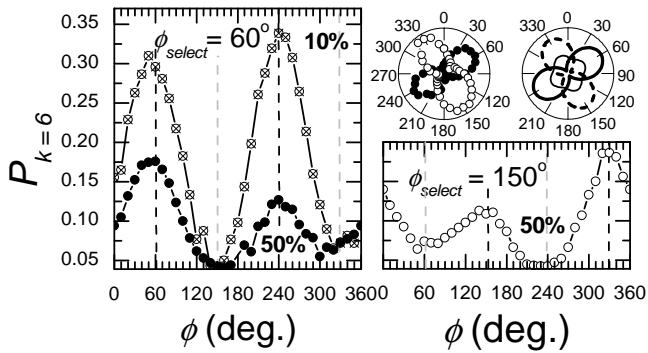}
\vspace{-5 mm}\caption{MQT relaxation probability versus $\phi$ for resonance $k$ = 6 for different selections of
the initial state of the sample. In the left figure the STF was applied along $\phi_{select}=60^o$, for different
initial states (50$\%$ and 10$\%$ of the biggest splittings of the distribution). The right figure shows the
result for a preparation of an initial state with 50$\%$ of the distribution with the orthogonal STF ,
$\phi_{select}=150^o$. The left polar plot shows the results for the biggest splittings (10$\%$ of the
distribution) for both selection angles. The polar plot on the right represents the calculated behavior of the
splitting for $E=+10$ mK (thick solid line), $E=-10$ mK (thick dashed line), $E=0$ (thin solid line).}\vspace{-8
mm}
\end{center}
\end{figure}

The observation of a two-fold symmetry in the MQT relaxation probability, with a 90$^o$ phase difference that
depends on the direction of the STF, is clear evidence that molecules in the crystal have lower than fourfold
symmetry. These observations are in excellent accord with the isomer model of Cornia et al. \cite{Cornia}. There
are six isomers in this model, four of which have lower symmetry than tetragonal (and comprise about 90$\%$ of the
molecules in the crystal) and thus have a second-order transverse magnetic anisotropy, $E(S^2_x-S^2_y)$. This
anisotropy gives rise to a two-fold rotation symmetry for fields in the $x$-$y$ plane, with maxima in the MQT
probability separated by 180$^o$. The maxima occur when the applied field is aligned with the medium magnetic axis
(for $E<$ 0, the $x$-axis and for $E>$ 0, the $y$-axis). Since the crystal has tetragonal symmetry there must be
equal populations of isomers with opposite signs of $E$. A change in the sign of $E$ rotates the hard and medium
axes of the molecule by 90 degrees, explaining the average four-fold symmetry observed in the crystal (fig. 1).
Moreover, the isomers found in x-ray diffraction have their hard axes in the following directions, 50-60$^o$,
140-150$^o$, 230-240$^o$, and 320-330$^o$ with respect to crystal faces \cite{Sessoli}, as we observe in Fig. 2.
Importantly, this observation is not consistent with the dislocation model \cite{Chudnovsky}, since in this model
the distribution of medium axes directions is isotropic, excluding very small regions of the crystal near the
dislocation cores. The polar plot on the right in fig. 2 shows the calculated tunnel splittings for resonance $k$
= 6 versus $\phi$. The solid thin line represents the angular dependence of the tunnel splitting for $E$ = 0. An
estimation of the magnitudes of $E$ needed to explain the change in the MQT relaxation observed in figs. 1 and 2
are: $E \sim$ 0.5 mK (fig. 1 upper curve, molecules with the smallest tunnel splittings), $E \sim$ 2.5 mK (fig. 1
lower curve, the whole distribution), $E \sim$ 2.5 mK (fig. 2, 50$\%$ of the largest tunnel splittings with
transverse field selection) and $E \sim$ 10 mK (fig. 2, 10$\%$ with the same transverse field selection). Recent
EPR experiments suggest similar values of $E$ \cite{Hill2}.

By applying a STF in a given direction we can select molecules with a particular direction of their medium axes.
We have thus selected molecules with their medium axis along $\phi_{select}=60^o$, and 10$\%$ and 50$\%$ of the
distribution of such molecules, for further investigation (as in fig. 2 on left). We have studied the MQT
relaxation as a function of the magnitude of the transverse field applied along both the hard ($\phi=150^0$) and
medium ($\phi=60^o$) axes of such molecules. The results for resonances $k=$ 5, 6, and 7 are shown in fig. 3. The
probability is plotted on a logarithmic scale versus the transverse magnetic field. For a given resonance, the
difference between these orientations of the transverse field is clearly larger in the experiment with 10$\%$ of
the molecules (biggest splittings) than with 50$\%$. This indicates that the average value of the second order
transverse anisotropy increases as we select the largest tunnel splittings in the distribution. An estimation
gives average $E$ values of $E(50\%)\sim 2.5mK$, $E(10\%)\sim 10mK$. These are the same values as those extracted
from the change of the probability in the rotation experiments.
\begin{figure}
\begin{center}\includegraphics[width=7.5cm]{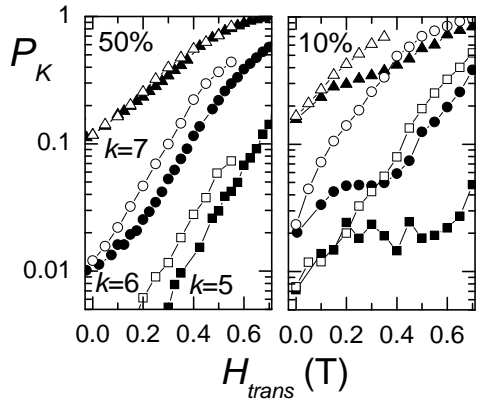}
\vspace{-3 mm}\caption{Transverse field dependence of the MQT relaxation probability for resonances $k$ = 5, 6 and
7. The transverse field is applied along the hard axis (solid symbols) and along the medium axis (open
symbols).}\vspace{-5 mm}
\end{center}
\end{figure}
One significant difference in the transverse field dependence of the probability can be observed through the
comparison of the graphics in fig. 3. In the results for 50$\%$ of the distribution, the probability increases
exponentially for both transverse field orientations. However, in the experiment with 10$\%$ of the molecules the
MQT relaxation probability deviates significantly from exponential behavior, showing the largest deviations at
fields $H_{p}(k=5)\sim$ 0.45 T, $H_{p}(k=6)\sim$ 0.3 T and $H_{p}(k=7)\sim$ 0.35 T. This is reminiscent of the
Berry phase observed in the SMM Fe$_8$ \cite{Wernsdorfer2}. The modeling that we have carried out using the values
of $E$ extracted from our data give minima in the tunneling probability for $H_{min}(k=5)\sim$ 0 and 0.5 T,
$H_{min}(k=6)\sim$ 0.25 T and $H_{min}(k=7)\sim$ 0 and 0.6 T. Some of these minima occur where we observe plateaus
in fig. 3. The distribution of $E$ values in the experiment probably smoothes the minima leading to the plateaus
we observe. On the other hand, the measured MQT probability does not show parity effects for the resonances at
zero transverse field. In the absence of transverse fields, the symmetry breaking terms in the Hamiltonian are due
to the transverse anisotropy of the molecules. Fourth order transverse anisotropy only allows tunneling
transitions for resonances $k$ that are a multiple of 4 ($k=4i$), while second order anisotropy only allows
transitions for $k$ a multiple of 2 ($k=2i$). The observation of tunneling relaxation at odd resonances must be
due to the presence of local transverse fields. In the Cornia model this is explained naturally as small tilts of
the molecular axis of the isomers. Other possibilities include transverse dipolar and nuclear fields or
dislocations.

In summary, the average four-fold symmetry of MQT in Mn$_{12}$-acetate is due to equal populations of isomers with
biaxial anisotropy that have orthogonal medium magnetic axes. Measurements of the tunneling probability versus
transverse field show evidence for Berry phase effects probably due to a combination of second and fourth order
transverse magnetic anisotropies. From a broader perspective, these results illustrate how subtle changes in
molecule environment can modulate magnetic anisotropy and magnetic quantum tunneling. This sensitivity to local
chemical environments is likely to be general to magnetism in molecules and ultimately useful in controlling their
quantum properties.

Acknowledgments. The authors grateful acknowledge useful discussions with S. Hill. This research was supported by
NSF (DMR 0103290). E. del B. acknowledges support from S.E.E.U. of Spain and Fondo Social Europeo.

$^*$Email address: delbarco@physics.nyu.edu\\ $^\dag$Email address: andy.kent@nyu.edu

\end{document}